%% file: main_cyue.tex
\documentclass[conference]{IEEEtran}
\IEEEoverridecommandlockouts 
\usepackage[T1]{fontenc}
\usepackage{cite}
\usepackage{color}
\usepackage[dvipsnames]{xcolor}

\usepackage[caption=false,font=footnotesize]{subfig}
\usepackage{amsmath,amssymb,bm}
\usepackage{graphicx}
\usepackage{booktabs,array}
\usepackage{url}

\usepackage{tikz}
\usepackage{pgfplots}
\pgfplotsset{compat=1.18}
\usetikzlibrary{arrows.meta,calc,shapes.geometric}
\input{figures/tikz/plotstyle}
\usepackage[hidelinks]{hyperref}

\graphicspath{{figures/}}
\makeatletter
\def\section{\@startsection{section}{1}{\z@}{2pt}{1pt}%
  {\normalfont\normalsize\centering\scshape}}
\def\subsection{\@startsection{subsection}{2}{\z@}{2pt}{0.5pt}%
  {\normalfont\normalsize\itshape}}
\makeatother
\IEEEaftertitletext{\vspace{-0.5\baselineskip}}
\begin{document}
\title{Semantic-Assisted Joint Channel Estimation and Symbol Detection for MU-MIMO Systems
\thanks{The work of Chentao Yue is supported by the ARC DECRA under Grant DE250101332. Code available: https://github.com/HaoChang1/Sem-JCED}}
\author{\IEEEauthorblockN{Hao Chang, Chentao Yue, Haiyao Yu, Yonghui Li, Branka Vucetic, and Wibowo Hardjawana}
\IEEEauthorblockA{School of Electrical and Computer Engineering, The University of Sydney, Australia\\
\{hao.chang, chentao.yue, haiyao.yu, yonghui.li, branka.vucetic, wibowo.hardjawana\}@sydney.edu.au}}
\maketitle

\begin{abstract}
This paper presents a semantic-assisted joint channel estimation and symbol detection (Sem-JCED) receiver framework for text transmission in multi-user multiple-input multiple-output (MU-MIMO) systems. Sem-JCED integrates a linear minimum mean-square error (LMMSE) channel estimator, an expectation propagation (EP) symbol detector, and a semantic corrector based on ByT5, a byte-level language model. At each outer iteration, LMMSE estimates the channel from known pilots and data symbol feedback, and EP detects the data symbols. ByT5 then maps the data symbol probabilities to byte probabilities, which provide soft symbol statistics and semantic priors for feedback to LMMSE and EP, respectively, in the next iteration.
Simulation results show that Sem-JCED requires about 7~dB lower signal-to-noise ratio (SNR) than conventional JCED to reach a bit error rate (BER) of $1\times10^{-2}$ with four users and four receive antennas, and outperforms maximum-likelihood detection with perfect channel knowledge over the simulated SNR range.

\end{abstract}
\vspace{9pt}
\section{Introduction}
Multi-user multiple-input multiple-output (MU-MIMO) systems support simultaneous transmissions from multiple users over shared time-frequency resources. This improves spectral efficiency but introduces multi-user interference (MUI). Reliable communication relies on symbol detection to mitigate MUI using accurate channel state information (CSI) obtained through channel estimation.
A common receiver architecture performs these tasks sequentially. The CSI is first estimated from known pilots, for example, by a linear minimum mean-square error (LMMSE) estimator~\cite{edfors1998}, and then passed to a detector, such as maximum-likelihood (ML) or expectation propagation (EP)~\cite{cespedes2014}. In this conventional design, channel estimation errors affect detection, while the detected data are not used to refine the initial channel estimates.

Joint channel estimation and symbol detection (JCED)~\cite{park2015virtual,11087618,10907797,11153399} introduces feedback between these two tasks. In a typical iterative JCED receiver, detected symbols and their reliability information provide additional observations for channel refinement, and the updated channel estimates improve symbol detection in the next iteration. For example, \cite{park2015virtual,11087618} use soft symbol estimates from the detector as virtual pilots for channel estimation in MIMO systems. 
Recently, \cite{10907797,11153399} developed neural-network-based JCED receivers that improve both channel estimation and symbol detection. However, the performance of JCED depends on the reliability of the feedback, because incorrect data estimates can contaminate the channel refinement and reinforce errors in subsequent iterations.

Pretrained large language models (LLMs) capture contextual dependencies in natural language, providing source information that helps recover corrupted text. This capability has motivated their integration with conventional receivers to support channel decoding~\cite{11432683,yue2026secc,yue2026osd,li2026viterbi,wei2026semir} and data-aided channel estimation~\cite{park2026semanticpilot,park2026}. The semantic error control coding framework incorporates source information from LLMs into established decoding algorithms~\cite{11432683,yue2026secc}. One such model is ByT5, a pretrained byte-level Transformer that operates directly on text bytes~\cite{xue2022byt5}. Its encoder can be fine-tuned with a byte-classification head to predict original-byte probabilities from corrupted text, providing semantic reliability information to assist conventional
decoding~\cite{yue2026osd,wei2026semir}. Semantic ordered-statistics decoding uses these probabilities to guide reliability ordering and candidate search~\cite{yue2026osd}, while LLM-Viterbi incorporates LLM information into convolutional-code path evaluation~\cite{li2026viterbi}. Extending semantic assistance to iterative multi-user reception, Sem-IR integrates ByT5 with interference cancellation and low-density parity-check decoding for users transmitting segments of a shared sentence over an additive white Gaussian noise channel~\cite{wei2026semir}. 
Semantic assistance also supports data-aided channel estimation, where LLM outputs help select reliable virtual pilots for single-input single-output systems~\cite{park2026semanticpilot}. Its single-user MIMO extension~\cite{park2026} further admits LLM-corrected symbols as virtual pilots when they pass a residual test on the received signal. In both works, the LLM output enters only the channel estimator, as hard symbol decisions.
However, these studies do not feed soft semantic information into both channel estimation and symbol detection within an iterative JCED framework for MU-MIMO systems.

In this paper, we propose a semantic-assisted JCED (Sem-JCED) receiver for MU-MIMO text transmission. Sem-JCED iterates between an LMMSE channel estimator, an EP symbol detector, and a ByT5 semantic corrector. 
The symbol probabilities estimated by EP are passed to ByT5, whose output byte probabilities are used to derive soft symbol
statistics and semantic priors for feedback to LMMSE and EP, respectively, so that semantic information assists channel estimation and symbol detection.
Simulations show that the normalized mean-square error (NMSE) of Sem-JCED approaches that of genie-aided LMMSE as the signal-to-noise ratio (SNR) increases and that, with four users and four receive antennas, Sem-JCED reaches a bit error rate (BER) of $1\times10^{-2}$ at about 7~dB lower SNR than conventional JCED. 

\textbf{Notation:} $a$, $\mathbf{a}$, and $\mathbf{A}$ denote a scalar, vector, and matrix, respectively.
$\mathbf{A}^{\top}$ and $\mathbf{A}^{H}$ denote the transpose and conjugate transpose of matrix $\mathbf{A}$, respectively.
$\mathbb{C}^{M\times N}$ and $\mathbb{R}^{M\times N}$ denote the spaces of $M\times N$ complex-valued and real-valued matrices, respectively. $\mathbf{I}_M$ denotes an $M\times M$ identity matrix.
$\|\cdot\|_2$ denotes the Euclidean norm.
$\mathbf{x}\sim\mathcal{CN}(\boldsymbol{\mu},\boldsymbol{\Sigma})$ means that $\mathbf{x}$ follows the circularly symmetric complex Gaussian distribution with mean $\boldsymbol{\mu}$ and covariance $\boldsymbol{\Sigma}$.
$\operatorname{diag}(\cdot)$ forms a diagonal matrix from its arguments, $\mathbb{E}[\cdot]$ denotes expectation, and $|\mathcal{A}|$ denotes the cardinality of a set $\mathcal{A}$.
The $(i,j)$th entry of $\mathbf{A}$ is denoted by $\mathbf{A}[i,j]$.

\section{System Model and Problem Formulation}
\subsection{System Model}
\label{sec:system_model}
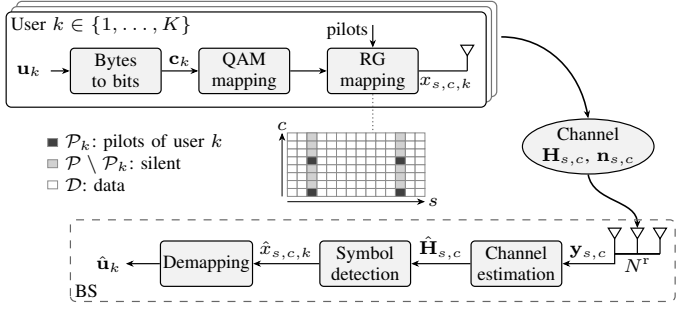
\begin{figure}[!t]
 \centering
 \input{figures/tikz/fig_system}%
 \par\vspace{-8pt}
 \caption{MU-MIMO system model, with a schematic RG of user $k$.}
 \label{fig:system}
\end{figure}
We consider an uplink MU-MIMO system with orthogonal frequency division multiplexing (OFDM) for uncompressed text transmission, in which a base station (BS) equipped with $N^{\rm r}$ receive antennas simultaneously serves $K$ single-antenna users. We do not consider channel coding in this paper. The gains in channel estimation and symbol detection are therefore attributable to the semantic prior rather than to coding redundancy. The overall system model is shown in Fig.~\ref{fig:system}. 
For the $k$-th user, $k\in\{1,\dots,K\}$, the text payload is represented as a byte sequence $\mathbf{u}_k\in\{0,\dots,255\}^{N^{\rm b}}$ of length $N^{\rm b}$ and converted into the information bit vector $\mathbf{c}_k\in\{0,1\}^{8N^{\rm b}}$. An $M$-ary quadrature amplitude modulation ($M$-QAM) scheme maps each group of $Q\!=\!\log_2 M$ information bits to a complex-valued data symbol. The $M$-QAM constellation set is denoted by $\boldsymbol{\Omega}$ and has unit average power. The data and pilot symbols are then mapped to the OFDM resource grid (RG).

The OFDM RG contains $N^{\rm sc}$ subcarriers and $N^{\rm sym}$ OFDM symbols. Each resource element (RE) is indexed by $(s,c)\in\mathcal{G}=\{1,\dots,N^{\rm sym}\}\times\{1,\dots,N^{\rm sc}\}$, where $s$ and $c$ denote the OFDM symbol and subcarrier indices, respectively.
For user $k$, let $\mathcal{P}_k$ denote the set of RE index pairs reserved for its pilots, and let $\mathcal{P}=\bigcup^{K}_{k=1}\mathcal{P}_k$ collect the pilot indices of all users. Different users place their pilots on different subcarriers, and user $k$ transmits nothing on $\mathcal{P}\setminus\mathcal{P}_k$, so that user $k$'s pilots do not interfere with other users.
We use $\mathcal{D}$ to represent the set of RE index pairs assigned to data symbols. The pilot and data sets satisfy $\mathcal{P}\cap\mathcal{D}=\emptyset$ and $\mathcal{P}\cup\mathcal{D}=\mathcal{G}$. For each user, $N^{\rm rg}=N^{\rm sc}N^{\rm sym}$, $N^{\rm d}=|\mathcal{D}|$, and $N^{\rm p}=|\mathcal{P}_k|$ denote the total number of REs and the numbers of data and pilot REs, respectively. These numbers are identical for all users. The $8N^{\rm b}$ bits of each user fill its $N^{\rm d}$ data REs, i.e., $QN^{\rm d}=8N^{\rm b}$.

For $(s,c)\in\mathcal{G}$, the received signal at the BS is modeled as
\begin{equation}
\begin{aligned}
    \mathbf{y}_{s,c}
    &= \mathbf{H}_{s,c} \mathbf{x}_{s,c}    +    \mathbf{n}_{s,c},
\end{aligned}
    \label{eq:complex_re_model}
\end{equation}
where $\mathbf{y}_{s,c}=[y_{1,s,c},\ldots,y_{r,s,c},\ldots,y_{N^{\rm r},s,c}]^{\top}\in\mathbb{C}^{N^{\rm r}}$ is the received signal, $\mathbf{H}_{s,c}\in\mathbb{C}^{N^{\rm r}\times K}$ is the channel matrix, and $\mathbf{x}_{s,c}=[x_{s,c,1},\dots,x_{s,c,k},\dots,x_{s,c,K}]^{\top}\in\mathbb{C}^{K}$ collects the transmitted symbols of all $K$ users on RE $(s,c)$. The noise vector is modeled as $\mathbf{n}_{s,c}\sim\mathcal{CN}(\mathbf{0},\sigma^2\mathbf{I}_{N^{\rm r}})$, and $\sigma^2$ is assumed perfectly known at the receiver.
Let $h_{r,k,s,c}=\mathbf{H}_{s,c}[r,k]$ denote the channel coefficient from user $k$ to antenna $r$ on RE $(s,c)$, and let $\mathbf{h}_{r,k}\in\mathbb{C}^{N^{\rm rg}}$ collect $\{h_{r,k,s,c}\}_{(s,c)\in\mathcal{G}}$ in a fixed RE order. Channel estimation operates on $\mathbf{h}_{r,k}$ for each antenna-user pair $(r,k)$, whereas symbol detection operates on $\mathbf{H}_{s,c}$ for each RE $(s,c)$.

\subsection{Problem Formulation}
\label{sec:problem_formulation}
For each $(s,c)\in\mathcal{D}$, the receiver aims to recover $\mathbf{x}_{s,c}$ from $\mathbf{y}_{s,c}$ in \eqref{eq:complex_re_model}. In the ideal case where the ground-truth channel matrix $\mathbf{H}_{s,c}$ is available, the maximum a posteriori (MAP) detector reduces to the maximum-likelihood (ML) detector when all symbols in the constellation set are equiprobable. The corresponding symbol detection problem for $(s,c)\in\mathcal{D}$ can be written as
\begin{equation}
    \mathbf{\hat{x}}_{s,c}
    =
    \operatorname*{argmin}_{\mathbf{x}_{s,c}\in \boldsymbol{\Omega}^{K}} \|\mathbf{y}_{s,c} - \mathbf{H}_{s,c} \mathbf{x}_{s,c}\|^2_2.
    \label{eq:ml}
\end{equation}
For text, however, the transmitted bytes, and hence the symbols, are neither equiprobable nor independent across REs, so \eqref{eq:ml} ignores the source redundancy.
Moreover, the receiver must estimate the unknown channel matrices $\{\mathbf{H}_{s,c}\}_{(s,c) \in \mathcal{D}}$ before detecting the transmitted data symbols $\{\mathbf{x}_{s,c}\}_{(s,c) \in \mathcal{D}}$. Pilot observations $\{\mathbf{y}_{s,c}\}_{(s,c) \in \mathcal{P}}$ provide initial channel estimates that are used for symbol detection. The resulting soft symbol estimates, together with the received data observations, can then support channel refinement, forming the conventional iterative JCED loop~\cite{park2015virtual}. However, inaccurate channel estimates can impair symbol detection, while erroneous detector feedback can propagate through subsequent channel updates. The contextual structure of each user's text provides additional information that can help resolve residual detection ambiguities. This motivates the proposed Sem-JCED receiver, which incorporates semantic information into both channel estimation and symbol detection.

\section{Proposed Scheme}\label{sec:proposed}
\begin{figure}[!t]
 \centering
 \input{figures/tikz/fig_receiver}%
 \par\vspace{-8pt}
 \caption{Proposed Sem-JCED receiver. Dashed arrows carry the feedback used in outer iteration $\ell+1$.}
 \label{fig:proposed}
\end{figure}
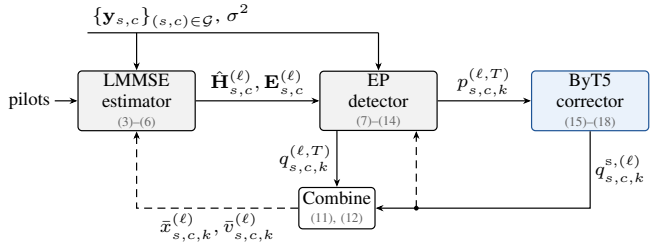

Sem-JCED comprises an LMMSE channel estimator, an EP detector, and a ByT5 semantic corrector, as shown in Fig.~\ref{fig:proposed}.
The modules run for outer iterations $\ell\in\{0,\dots,L\}$, where $\ell=0$ uses only the pilots for LMMSE and a uniform symbol prior for EP. In outer iteration $\ell$, the LMMSE channel estimator produces the estimated channel matrix $\hat{\mathbf{H}}_{s,c}^{(\ell)}$ and error variances $\mathbf{E}^{(\ell)}_{s,c}$ based on the known pilots and feedback data symbol estimates.
EP then uses the estimated channel to compute the symbol probabilities $p^{(\ell,T)}_{s,c,k}$. 
ByT5 takes the hard bit decisions on $p^{(\ell,T)}_{s,c,k}$ as input and produces the semantic prior $q^{\rm s,(\ell)}_{s,c,k}$, which is combined with the EP cavity distribution to obtain the data symbol means and variances $\bar x_{s,c,k}^{(\ell)}, \bar v_{s,c,k}^{(\ell)}$. The means and variances are fed back to the LMMSE channel estimator, and $q^{\rm s,(\ell)}_{s,c,k}$ to the EP detector.

\subsection{LMMSE Channel Estimation}\label{sec:lmmse_ce}
At the $\ell$-th iteration, the inputs of LMMSE include the received signals $\{\mathbf y_{s,c}\}_{(s,c)\in\mathcal G}$,
the noise variance $\sigma^2$, the known pilot symbols $x_{s,c,k}$, $(s,c)\in\mathcal P_k$, and, for $\ell\geq1$, the feedback data symbol means and variances $\{\bar x_{s,c,k}^{(\ell-1)}, \bar v_{s,c,k}^{(\ell-1)}\}_{(s,c)\in\mathcal D}$ from the previous iteration, which are obtained by combining the ByT5 semantic priors with the EP cavity distributions, as described in Section~\ref{sec:byt5_feedback}.
For each receive antenna $r$ and user $k$, LMMSE provides the channel estimate $\hat{\mathbf h}_{r,k}^{(\ell)}\in\mathbb C^{N^{\rm rg}}$ and the associated error variance matrix $\mathbf E_{r,k}^{(\ell)} \in\mathbb C^{N^{\rm rg}\times N^{\rm rg}}$.

At $\ell=0$, the known pilots provide channel observations on $(s,c)\in\mathcal{P}_k$, given as
\begin{equation}\label{eq:ce_pilot_obs}
 \widetilde h_{r,k,s,c}^{(0)}
 =
 \frac{y_{r,s,c}}{x_{s,c,k}},
 \qquad
 w_{r,k,s,c}^{(0)}
 =
 \frac{|x_{s,c,k}|^2}{\sigma^2},
\end{equation}
where $\widetilde h_{r,k,s,c}^{(0)}$ is the channel observation and $w_{r,k,s,c}^{(0)}$ is the corresponding precision weight for user $k$, antenna $r$, and RE $(s,c)$.
Let $\widetilde{\mathbf h}_{r,k}^{(\ell)} \in\mathbb C^{N^{\rm rg}}$ collect the channel observations for all REs, and let $\mathbf W_{r,k}^{(\ell)} \in\mathbb R^{N^{\rm rg}\times N^{\rm rg}}$ be the diagonal matrix of their precision weights.
For $\ell=0$, both channel observations and weights outside $\mathcal P_k$ are zero.

For $\ell\geq1$, the pilot entries are retained, i.e., $\widetilde h_{r,k,s,c}^{(\ell)}=\widetilde h_{r,k,s,c}^{(0)}$ and $w_{r,k,s,c}^{(\ell)}=w_{r,k,s,c}^{(0)}$ for $(s,c)\in\mathcal{P}_k$, and the entries on $\mathcal{P}\setminus\mathcal{P}_k$ remain zero, while the data entries are updated using the feedback symbol means $\bar x_{s,c,k}^{(\ell-1)}$ and variances $\bar v_{s,c,k}^{(\ell-1)}$.
For a data RE $(s,c)\in\mathcal D$, soft interference cancellation constructs the virtual channel observation and its precision weight as
\begin{subequations}\label{eq:virtual}
\begin{align}
 \widetilde h_{r,k,s,c}^{(\ell)}
 &=
 \frac{
 y_{r,s,c}
 -\sum^K_{i=1,i\ne k}
 \hat h_{r,i,s,c}^{(\ell-1)}
 \bar x_{s,c,i}^{(\ell-1)}
 }{
 \bar x_{s,c,k}^{(\ell-1)}
 },\\
 w_{r,k,s,c}^{(\ell)}
 &=
 \frac{
 \left|\bar x_{s,c,k}^{(\ell-1)}\right|^2
 }{
 \sigma^2+
 \sum_{i=1}^{K}
 \left|\hat h_{r,i,s,c}^{(\ell-1)}\right|^2
 \bar v_{s,c,i}^{(\ell-1)}
 }.
\end{align}
\end{subequations}
Here, $\hat h_{r,k,s,c}^{(\ell-1)}$ is the estimated coefficient associated with RE $(s,c)$ in the previous iteration. The precision weight accounts for noise and residual symbol uncertainty.
 
The LMMSE update is then given as
\begin{subequations}\label{eq:softlmmse}
\begin{align}
 \mathbf E_{r,k}^{(\ell)}
 &=
 \left(
 \mathbf I_{N^{\rm rg}}
 +\mathbf R_{r,k}\mathbf W_{r,k}^{(\ell)}
 \right)^{-1}
 \mathbf R_{r,k},\\
 \hat{\mathbf h}_{r,k}^{(\ell)}
 &=
 \mathbf E_{r,k}^{(\ell)}
 \mathbf W_{r,k}^{(\ell)}
 \widetilde{\mathbf h}_{r,k}^{(\ell)},
\end{align}
\end{subequations}
where $\mathbf R_{r,k}=\mathbb E[\mathbf h_{r,k}\mathbf h_{r,k}^{H}]$ is computed from the known statistical parameters of the adopted channel model~\cite{edfors1998}, and the expectation is taken over random channel realizations.

We then smooth the estimates for $\ell\geq1$ by damping,
\begin{subequations}\label{eq:ce_damping}
\begin{align}
 \hat{\mathbf h}_{r,k}^{(\ell)}
 &\leftarrow
 (1-\eta^{\rm lmmse})
 \hat{\mathbf h}_{r,k}^{(\ell)}
 +\eta^{\rm lmmse}
 \hat{\mathbf h}_{r,k}^{(\ell-1)},\\
 \mathbf E_{r,k}^{(\ell)}
 &\leftarrow
 (1-\eta^{\rm lmmse})
 \mathbf E_{r,k}^{(\ell)}
 +\eta^{\rm lmmse}
 \mathbf E_{r,k}^{(\ell-1)},
\end{align}
\end{subequations}
where $\eta^{\rm lmmse}$ weights the previous and current estimates.
Let $E_{r,k,s,c}^{(\ell)}$ denote the diagonal entry of $\mathbf E_{r,k}^{(\ell)}$ associated with RE $(s,c)$, i.e., the error variance of $\hat h_{r,k,s,c}^{(\ell)}$. For each data RE, we form $\hat{\mathbf H}_{s,c}^{(\ell)} \in\mathbb C^{N^{\rm r}\times K}$ and $\mathbf E_{s,c}^{(\ell)} \in\mathbb R^{N^{\rm r}\times K}$ with $ \hat{\mathbf H}_{s,c}^{(\ell)}[{r,k}] = \hat h_{r,k,s,c}^{(\ell)}$ and $ {\mathbf E}_{s,c}^{(\ell)}[{r,k}] = E_{r,k,s,c}^{(\ell)}$.
These two estimates for $(s,c)\in\mathcal{D}$ are forwarded to the EP symbol detector, as shown in Fig.~\ref{fig:proposed}.

\subsection{Expectation Propagation Symbol Detection}\label{sec:ep_sd}
EP performs symbol detection for all users on each data RE $(s,c)\in\mathcal D$ by iteratively refining Gaussian site functions through observation and estimation modules~\cite{cespedes2014}.
At outer iteration $\ell$, its inputs are the received signal $\mathbf y_{s,c}$, the noise variance $\sigma^2$, the channel estimate $\hat{\mathbf H}_{s,c}^{(\ell)}$, the error variances $\mathbf E_{s,c}^{(\ell)}$, and the semantic symbol priors $q_{s,c,k}^{\rm s,(\ell-1)}(a)$ obtained from ByT5 in iteration $\ell-1$.
After $T$ inner iterations, it outputs the posterior probabilities $p_{s,c,k}^{(\ell,T)}(a)$ and the cavity distributions $q_{s,c,k}^{(\ell,T)}(a)$, where $a\in\boldsymbol\Omega$ denotes a constellation point.

For each data RE $(s,c)$, treating the channel estimation error as additional noise, we approximate the effective noise covariance as
\begin{equation}\label{eq:whitening}
 \mathbf S_{s,c}^{(\ell)}
 =
 \sigma^2\mathbf I_{N^{\rm r}}
 +
 \operatorname{diag}\!\left(
 \sum_{k=1}^{K}E_{1,k,s,c}^{(\ell)},
 \ldots,
 \sum_{k=1}^{K}E_{N^{\rm r},k,s,c}^{(\ell)}
 \right).
\end{equation}

The following EP updates apply to an arbitrary data RE. For notational simplicity, we omit $(s,c)$ and write
$\mathbf y$, $\hat{\mathbf H}^{(\ell)}$, and $\mathbf S^{(\ell)}$ for the corresponding RE quantities.
We retain the outer iteration index $\ell$ and the inner iteration index $t=1,\ldots,T$.
The channel estimate, effective noise covariance, and semantic priors remain fixed throughout the inner iterations.

\subsubsection{Observation}
At inner iteration $t$, the mean $\boldsymbol\mu^{(\ell,t)}\in\mathbb C^K$ and covariance $\boldsymbol\Sigma^{(\ell,t)}\in\mathbb C^{K\times K}$ of the Gaussian approximation are given as
\begin{subequations}\label{eq:epgauss}
\begin{align}
 \boldsymbol\Sigma^{(\ell,t)}
 &=\Bigl(
 (\hat{\mathbf H}^{(\ell)})^H
 (\mathbf S^{(\ell)})^{-1}
 \hat{\mathbf H}^{(\ell)}
 +\boldsymbol\Lambda^{(\ell,t-1)}
 \Bigr)^{-1},\\
 \boldsymbol\mu^{(\ell,t)}
 &=\boldsymbol\Sigma^{(\ell,t)}
 \Bigl(
 (\hat{\mathbf H}^{(\ell)})^H
 (\mathbf S^{(\ell)})^{-1}\mathbf y
 +\boldsymbol\gamma^{(\ell,t-1)}
 \Bigr).
\end{align}
\end{subequations}
The diagonal matrix $\boldsymbol\Lambda^{(\ell,t)}=\mathrm{diag}(\lambda_{1}^{(\ell,t)},\dots,\lambda_{K}^{(\ell,t)})$ collects the site precisions, and $\boldsymbol\gamma^{(\ell,t)}=[\gamma_{1}^{(\ell,t)},\dots,\gamma_{K}^{(\ell,t)}]^{\top}$ collects the corresponding precision means.
They are initialized as $\lambda_k^{(\ell,0)}=1$ and $\gamma_k^{(\ell,0)}=0$.
We use $\mu_{k}^{(\ell,t)}$ for the $k$-th mean entry and $\Sigma_{k}^{(\ell,t)}$ for the $k$-th diagonal covariance entry.
Removing the $k$-th site gives the cavity distribution, whose density at $a$ is
\begin{equation}\label{eq:ep_cavity_density}
 q_{k}^{(\ell,t)}(a)
 =\frac{1}{\pi v_{{\rm obs},k}^{(\ell,t)}}\exp\!\left(-\frac{\bigl|a-x_{{\rm obs},k}^{(\ell,t)}\bigr|^2}{v_{{\rm obs},k}^{(\ell,t)}}\right),
\end{equation}
where
\begin{subequations}\label{eq:cavity}
\begin{align}
 v_{{\rm obs},k}^{(\ell,t)}
 &=\left(
 \frac{1}{\Sigma_{k}^{(\ell,t)}}
 -\lambda_{k}^{(\ell,t-1)}\right)^{-1},\\
 x_{{\rm obs},k}^{(\ell,t)}
 &=v_{{\rm obs},k}^{(\ell,t)}
 \left(
 \frac{\mu_{k}^{(\ell,t)}}{\Sigma_{k}^{(\ell,t)}}
 -\gamma_{k}^{(\ell,t-1)}\right).
\end{align}
\end{subequations}

\subsubsection{Estimation}
The cavity distribution $q_k^{(\ell,t)}(a)$ is combined with the semantic prior $q_k^{\rm s,(\ell-1)}(a)$ from the preceding outer iteration, obtained as described in Section~\ref{sec:byt5_feedback}.
The combination of the probabilities gives
\begin{equation}\label{eq:semantic_tilted}
 p_k^{(\ell,t)}(a)
 =
 \frac{
 q_k^{(\ell,t)}(a)q_k^{\rm s,(\ell-1)}(a)}
 {\sum_{a'\in\boldsymbol\Omega}
 q_k^{(\ell,t)}(a')q_k^{\rm s,(\ell-1)}(a')}.
\end{equation}
At $\ell=0$, no ByT5 feedback is available, so the uniform prior $q_k^{\rm s,(-1)}(a)=1/M$ for all $a\in\boldsymbol\Omega$ reduces this expression to the normalized cavity distribution over the constellation.

The combined probabilities are used to calculate the posterior symbol mean and variance as
\begin{subequations}\label{eq:ep_estimation}
\begin{align}
 \hat x_{k}^{(\ell,t)}
 &=\sum_{a\in\boldsymbol\Omega}a\,p_{k}^{(\ell,t)}(a),\\
 \hat v_{k}^{(\ell,t)}
 &=\sum_{a\in\boldsymbol\Omega}
 |a-\hat x_{k}^{(\ell,t)}|^2 p_{k}^{(\ell,t)}(a).
\end{align}
\end{subequations}
The updated site parameters are given as
\begin{subequations}\label{eq:sites}
\begin{align}
 \lambda_{k}^{(\ell,t)}
 &=\frac{1}{\hat v_{k}^{(\ell,t)}}
 -\frac{1}{v_{{\rm obs},k}^{(\ell,t)}},\\
 \gamma_{k}^{(\ell,t)}
 &=\frac{\hat x_{k}^{(\ell,t)}}{\hat v_{k}^{(\ell,t)}}
 -\frac{x_{{\rm obs},k}^{(\ell,t)}}
 {v_{{\rm obs},k}^{(\ell,t)}}.
\end{align}
\end{subequations}
We note that if $\lambda_{k}^{(\ell,t)}$ is negative, which is invalid for an inverse variance term, we set $\lambda_{k}^{(\ell,t)}=\lambda_{k}^{(\ell,t-1)}$ and $ \gamma_{k}^{(\ell,t)}= \gamma_{k}^{(\ell,t-1)}$~\cite{11555386}.
They are then damped as
\begin{subequations}\label{eq:ep_damping}
\begin{align}
 \lambda_{k}^{(\ell,t)}
 &\leftarrow(1-\eta^{\rm ep})\lambda_{k}^{(\ell,t)}
 +\eta^{\rm ep}\lambda_{k}^{(\ell,t-1)},\\
 \gamma_{k}^{(\ell,t)}
 &\leftarrow(1-\eta^{\rm ep})\gamma_{k}^{(\ell,t)}
 +\eta^{\rm ep}\gamma_{k}^{(\ell,t-1)}.
\end{align}
\end{subequations}
The updated site parameters are fed back to~\eqref{eq:epgauss} for the next inner iteration. After $T$ inner iterations, EP produces symbol probabilities $\{p^{(\ell,T)}_{s,c,k}\}_{(s,c)\in\mathcal{D}}$ for all $K$ users and forwards them to ByT5.

\subsection{ByT5 Text Correction and Feedback}\label{sec:byt5_feedback}
The input symbol probabilities $\{p^{(\ell,T)}_{s,c,k}\}_{(s,c)\in\mathcal{D}}$ are converted into hard bit decisions using the fixed QAM bit labels.
Each bit is assigned the value with the larger marginal probability. 
Following the fixed RG demapping order, the decisions from all data REs are collected into $\hat{\mathbf c}_k^{(\ell)} = \left[ \hat c_{k,1}^{(\ell)},\ldots,
 \hat c_{k,QN^{\rm d}}^{(\ell)} \right]^{\top}\!\in\!\{0,1\}^{QN^{\rm d}}.$
It is then regrouped into bytes and processed independently for each user by ByT5 at every outer iteration.
Each user's payload consists of $N^{\rm b}/N^{\rm seq}$ sequences of $N^{\rm seq}$ bytes, which ByT5 processes separately. In the following, we describe one sequence of user $k$, omit its sequence index, and use bit and byte indices local to that sequence. Every eight consecutive bits form a detected byte,
\begin{equation}\label{eq:detected_byte}
 \hat u_{k,j}^{(\ell)}
 =\sum_{i=1}^{8}2^{8-i}\hat c_{k,8(j-1)+i}^{(\ell)},
\end{equation}
where $j=1,\ldots,N^{\rm seq}$. Collecting all the byte estimates results in $ \hat{\mathbf u}_k^{(\ell)}
 =[\hat u_{k,1}^{(\ell)},\ldots,\hat u_{k,N^{\rm seq}}^{(\ell)}]^\top.$

\begin{figure}[!t]
 \centering
 \input{figures/tikz/fig_byt5}%
 \par\vspace{-8pt}
 \caption{ByT5 semantic corrector for one sequence with an erroneous byte at position $j$ (illustrative probabilities).}
 \label{fig:byt5}
\end{figure}
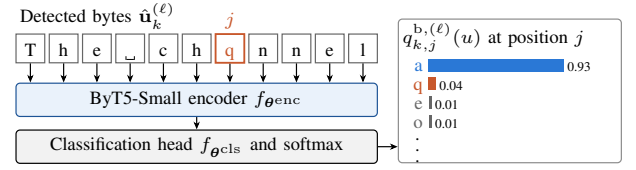

Following~\cite{wei2026semir}, we append a classification head with 256 outputs, one per byte value, to the pretrained ByT5-Small encoder~\cite{xue2022byt5}. The encoder hidden states are
\begin{equation}\label{eq:byt5_encoder} 
 \mathbf Z_k^{(\ell)}
 =f_{\boldsymbol\theta^{\rm enc}}
 (\hat{\mathbf u}_k^{(\ell)})\in\mathbb R^{N^{\rm seq}\times d},
\end{equation}
where $f_{\boldsymbol\theta^{\rm enc}}$ represents the ByT5 encoder with trainable parameters $\boldsymbol\theta^{\rm enc}$, and $d=1472$ is its hidden dimension.
Let $\mathbf z_{k,j}^{(\ell)}\in\mathbb R^d$ denote the $j$th row of $\mathbf Z_k^{(\ell)}$, i.e., the encoder hidden state at byte position $j$. The classification head with softmax activation produces
\begin{equation}\label{eq:byt5_output}
 \mathbf q_{k,j}^{\rm b,(\ell)}
 ={\rm{softmax}} (f_{\boldsymbol\theta^{\rm cls}}(\mathbf z_{k,j}^{(\ell)})),
\end{equation}
where $ \mathbf q_{k,j}^{\rm b,(\ell)}\in[0,1]^{256}$ collects the probabilities $q_{k,j}^{\rm b,(\ell)}(u)$ of the byte values $u\in\{0,\dots,255\}$ at position $j$, and $f_{\boldsymbol\theta^{\rm cls}}$ is the classification head with trainable parameters $\boldsymbol\theta^{\rm cls}$. As illustrated in Fig.~\ref{fig:byt5}, the context of the detected bytes lets ByT5 assign a high probability to the transmitted byte even at an erroneous position.
The encoder and the classification head are fine-tuned offline to recover the transmitted bytes from the detected bytes, as detailed in Section~\ref{sec:results}-A.

The ByT5 byte probabilities are converted into semantic symbol probabilities through the fixed bit-to-symbol
and RG mappings. For a fixed data RE, whose indices are omitted, its $Q$ bits belong to a single byte $j$, at bit positions $\mathcal{I}\subset\{1,\dots,8\}$, because $Q=4$ divides 8 for 16-QAM that is considered in this paper. Then
\begin{equation}\label{eq:byteqam}
 q_k^{\rm s,(\ell)}(a)
 =\sum_{u\in\mathcal{U}_{\mathcal{I}}(a)} q_{k,j}^{\rm b,(\ell)}(u),
\end{equation}
where $\mathcal{U}_{\mathcal{I}}(a)$ is the set of byte values whose bits at positions $\mathcal{I}$ equal the bit label of $a$.

For each data RE $(s,c)\in\mathcal D$, the semantic symbol probabilities $q_k^{\rm s,(\ell)}(a)$ are combined with the EP cavity distribution $q_k^{(\ell,T)}(a)$ as in~\eqref{eq:semantic_tilted}, i.e., $p_k^{\rm s,(\ell)}(a)\propto q_k^{(\ell,T)}(a)\,q_k^{\rm s,(\ell)}(a)$.
The resulting posterior $p_k^{\rm s,(\ell)}(a)$ is then used in~\eqref{eq:ep_estimation} to obtain the refined symbol mean $\bar x_{s,c,k}^{(\ell)}$ and variance $\bar v_{s,c,k}^{(\ell)}$.
These statistics provide the soft data estimates used to construct the virtual channel observations and
precision weights in~\eqref{eq:virtual} for the LMMSE channel estimation in the next outer iteration $\ell+1$.
The symbol probabilities $q_k^{\rm s,(\ell)}(a)$ are also supplied directly as the fixed semantic prior in~\eqref{eq:semantic_tilted} for EP in the next outer iteration $\ell+1$.
Since ByT5 takes hard decisions as input, an extrinsic form of $q_k^{\rm s,(\ell)}(a)$ is not directly available, and we use the ByT5 output as the prior.
After $L$ outer iterations, final hard bit decisions are obtained from the bit marginals of $p_k^{{\rm s},(L)}(a)$.

\begin{table}[t]
\caption{Simulation settings.}
\label{tab:settings}
\vspace{-5pt}
\centering\footnotesize
\begin{tabular}{@{}llll@{}}
\toprule
Parameter & Value & Parameter & Value\\
\midrule
$(K,N^{\rm r})$ & $(4,4)$, $(4,8)$ & Modulation & 16-QAM\\
Subcarriers & 128 & Subcarrier spacing & 30 kHz\\
OFDM symbols & 14 (2 pilot, 12 data) & Carrier frequency & 3.5 GHz\\
Channel model & TDL-C, 24 paths & Delay spread & 300 ns\\
\bottomrule
\end{tabular}
\end{table}

\section{Simulation Results}\label{sec:results}
\subsection{Simulation Settings}
Table~\ref{tab:settings} summarizes the OFDM settings with the tapped-delay-line C (TDL-C) channel model~\cite{3GPP38901}. Each OFDM frame carries $N^{\rm b}=768$ bytes per user, i.e., six sentences of $N^{\rm seq}=128$ bytes, in $N^{\rm d}=1536$ data REs. We denote a configuration by $K\times N^{\rm r}$.
The SNR is defined as $10\log_{10}(1/\sigma^2)$~dB. All users are stationary, with zero Doppler shift.

We fine-tune the ByT5-Small encoder with the classification head before deployment. We select 35,000 English sentences from Wikipedia and split them into 25,000, 5,000, and 5,000 sentences for training, validation, and testing, respectively. Each sentence contains 121--128 ASCII bytes and is zero-padded to 128 bytes. The padding bits are not counted in the BER.
The ByT5 inputs are the bytes detected by EP at $\ell=0$ when each sentence is transmitted in the $4\times4$ system at SNRs of 4, 8, 12, and 16~dB, which gives 140,000 samples in total. All simulation results are obtained with the 5,000 test sentences for each SNR.
Fine-tuning minimizes the cross-entropy between the byte probabilities in~\eqref{eq:byt5_output} and the transmitted bytes,
\begin{equation}\label{eq:ce_loss}
 \mathcal{L}\bigl(\boldsymbol\theta^{\rm enc},\boldsymbol\theta^{\rm cls}\bigr)
 =-\frac{1}{N^{\rm seq}}\sum_{j=1}^{N^{\rm seq}}\log q_{k,j}^{\rm b}(u_{k,j}),
\end{equation}
averaged over the training samples, where $u_{k,j}$ is the $j$th transmitted byte of the sequence.
We fine-tune ByT5 for 10 epochs, using the AdamW optimizer with a batch size of 32, a learning rate of $10^ {-4}$, and a weight decay of 0.01.
The number of EP inner iterations is $T=10$, and the damping factors are $\eta^{\rm lmmse}=0.5$ and $\eta^{\rm ep}=0.7$.

\subsection{Comparison Schemes}

To evaluate channel estimation and symbol detection performance, we consider the following schemes for comparison.
\begin{itemize}
    \item \textbf{Genie-aided LMMSE:}
    LMMSE estimation with all pilot and data symbols known, which serves as an NMSE benchmark for data-aided estimation.

    \item \textbf{LMMSE+EP:}
    Pilot-based LMMSE channel estimation~\cite{edfors1998}, i.e., LMMSE interpolation of the least-squares pilot estimates, is followed by EP detection~\cite{cespedes2014}, without iterative channel refinement.

    \item \textbf{Conv-JCED:}
    LMMSE channel estimation and EP detection are performed iteratively. Soft symbol means and variances from EP are used to construct the virtual pilots in~\eqref{eq:virtual} for channel refinement, similar to~\cite{11087618}.

    \item \textbf{Sem-JCED:}
    The proposed receiver, which uses ByT5 byte probabilities to construct virtual pilots for channel refinement and to modify the symbol probabilities within EP detection.

    \item \textbf{CSI-EP:}
    EP detection uses perfect CSI.

    \item \textbf{CSI-ML:}
    ML detection uses perfect CSI and exhaustively searches all $M^K$ symbol combinations.
\end{itemize}
The BER is the fraction of erroneous payload bits over all users and frames.
Channel NMSE is calculated over all evaluated frames and REs.

The LLM-aided estimator in~\cite{park2026} is not included as a baseline, because it is designed for a single-user $2\times2$ MIMO link with QPSK over a frequency-flat channel, and its fine-tuned generative LLM and selection threshold would have to be redesigned for our MU-MIMO-OFDM setting.

\subsection{Performance Evaluation}

\subsubsection{Convergence}
\begin{figure}[!t]
\centering
\input{figures/tikz/fig_convergence}%
\par\vspace{-3pt}
\caption{NMSE and BER versus the outer iteration $\ell$ for $K=N^{\rm r}=4$ at an SNR of 12~dB.}
\label{fig:convergence}
\end{figure}
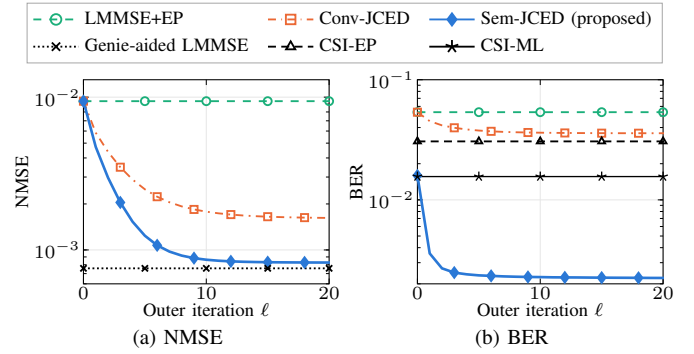

We first examine the convergence of the proposed Sem-JCED and the comparison schemes in Fig.~\ref{fig:convergence} for $K=N^{\rm r}=4$ at an SNR of 12~dB. The performance of channel estimation and symbol detection in terms of NMSE and BER is illustrated in Fig.~\ref{fig:convergence}(a) and Fig.~\ref{fig:convergence}(b), respectively. Both the NMSE and BER of Sem-JCED and Conv-JCED decrease with outer iteration $\ell$ and become nearly stable after approximately 10 iterations. 
We therefore use $L=10$ outer iterations for both Sem-JCED and Conv-JCED in the following comparisons.

\subsubsection{$4\times4$ system}
For the system with $K=N^{\rm r}=4$, Fig.~\ref{fig:snr}(a) shows that Sem-JCED outperforms both LMMSE+EP and Conv-JCED at all SNRs. Semantic corrections provide more reliable soft symbol estimates, which improve the virtual observations and their precision weights used for channel refinement. This is particularly beneficial when noisy detector outputs limit the accuracy of conventional data-aided estimation. As SNR increases, the detector outputs become more reliable, and both Conv-JCED and Sem-JCED gradually approach the genie-aided LMMSE.
Fig.~\ref{fig:snr}(b) shows that Sem-JCED requires about 7~dB lower SNR than Conv-JCED and CSI-EP to reach a BER of $1\times10^{-2}$. In addition to benefiting from refined channel estimates, it exploits the semantic prior in EP iterations. Sem-JCED even outperforms CSI-ML over the simulated SNR range, by around 4~dB at a BER of $1\times10^{-2}$ and 2~dB at $1\times10^{-3}$. Although CSI-ML uses perfect CSI, it does not exploit the contextual structure of the transmitted text, which provides Sem-JCED with additional information for recovering the payload bits.

\subsubsection{$4\times8$ system}
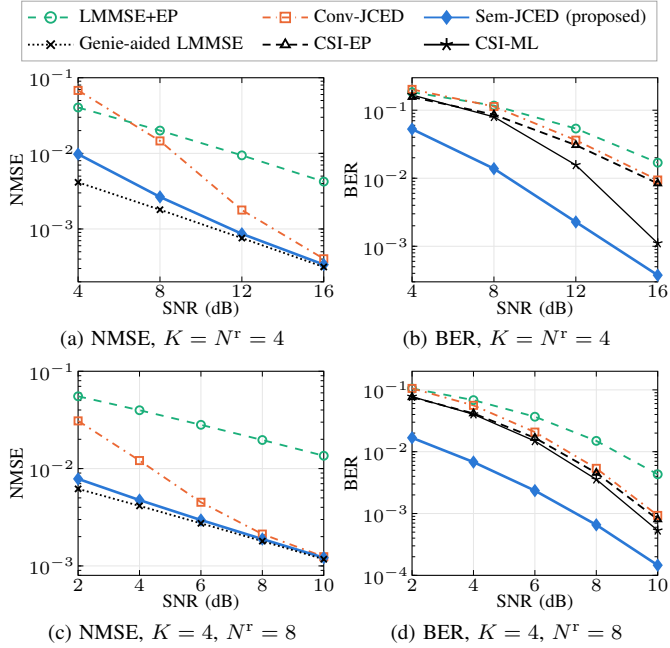
\begin{figure}[!t]
\centering
\input{figures/tikz/fig_snr}%
\par\vspace{-3pt}
\caption{NMSE and BER versus SNR.}
\label{fig:snr}
\end{figure}

We further evaluate the system with $K=4$ and $N^{\rm r}=8$. Consistent with the $4\times4$ results, Fig.~\ref{fig:snr}(c) shows that Sem-JCED improves channel estimation accuracy over LMMSE+EP and Conv-JCED at low SNRs and approaches the genie-aided LMMSE reference as SNR increases. In Fig.~\ref{fig:snr}(d), Sem-JCED outperforms Conv-JCED and CSI-EP by around 2.5 dB at a BER of $1\times10^{-3}$ and also outperforms CSI-ML by around 2 dB at a BER of $1\times10^{-3}$. 
Since ByT5 is fine-tuned only on the $4\times4$ system, these results show that its gains carry over to an antenna configuration not seen in training.

\begin{table}[t]
\caption{Runtime per OFDM frame (ms) for $K=N^{\rm r}=4$.}
\label{tab:latency}
\centering\footnotesize
\begin{tabular}{@{}lcccc@{}}
\toprule
Scheme & Channel estimation & Detection & ByT5 & Total\\
\midrule
LMMSE+EP & 1.89 & 18.29 & -- & 20.18\\
Conv-JCED & 33.54 & 179.62 & -- & 213.16\\
Sem-JCED & 34.19 & 187.41 & 235.29 & 456.89\\
CSI-EP & -- & 18.29 & -- & 18.29\\
CSI-ML & -- & 28.98 & -- & 28.98\\
\bottomrule
\end{tabular}
\end{table}

\subsection{Latency}
Table~\ref{tab:latency} lists the runtime per OFDM frame for $K=N^{\rm r}=4$, with $L=10$ for both JCED schemes. The 24 sentences from all users are processed in one ByT5 batch on an NVIDIA RTX 6000 Ada GPU.
Sem-JCED requires 456.89~ms per frame, compared with 213.16~ms for Conv-JCED, 20.18~ms for LMMSE+EP, 18.29~ms for CSI-EP, and 28.98~ms for CSI-ML.
ByT5 inference accounts for 235.29~ms, approximately 51.5\% of the total runtime.
Despite its higher latency, Sem-JCED achieves substantially better BER performance.
These results highlight the tradeoff between detection performance and processing latency.
Similar inference overhead has also been reported in other semantic-assisted communication systems~\cite{yue2026osd,wei2026semir}.

\section{Conclusion}
This paper proposed Sem-JCED, a semantic-assisted joint channel estimation and symbol detection framework for MU-MIMO-OFDM text transmission. The receiver combines LMMSE channel estimation, EP symbol detection, and ByT5 semantic correction. ByT5 byte probabilities provide soft symbol statistics for channel refinement and the semantic prior for EP detection.
Simulation results show that Sem-JCED requires about 7~dB lower SNR than Conv-JCED to reach a BER of $1\times10^{-2}$ in the $4\times4$ system, outperforms CSI ML over the simulated SNR range, and approaches the NMSE of genie-aided LMMSE estimation as the SNR increases.

\bibliographystyle{IEEEtran}
\bibliography{references}
\end{document}

%% file: figures/tikz/plotstyle.tex
\definecolor{cLMMSE}{HTML}{1BAF7A}
\definecolor{cConv}{HTML}{EB6834}
\definecolor{cSem}{HTML}{2A78D6}
\definecolor{cGrid}{HTML}{E6E6E6}

\pgfplotsset{
  resultaxis/.style={
    scale only axis,
    width=3.25cm, height=2.7cm,
    tick label style={font=\scriptsize, inner sep=1pt},
    label style={font=\scriptsize},
    xlabel style={yshift=3pt},
    ylabel style={inner sep=1pt},
    grid=major,
    major grid style={line width=0.25pt, draw=cGrid},
    axis line style={line width=0.4pt},
    major tick length=2pt, minor tick length=1.2pt,
    every tick/.style={line width=0.3pt, black},
    tick align=inside,
    mark size=1.5pt,
    every axis plot/.append style={line width=0.7pt},
  },
  sLMMSE/.style={color=cLMMSE, dashed, mark=o, mark options={solid, fill=white}},
  sConv/.style={color=cConv, dashdotted, mark=square, mark options={solid, fill=white, scale=0.9}},
  sSem/.style={color=cSem, solid, line width=1pt, mark=diamond*, mark options={solid, fill=cSem, scale=1.15}},
  sGenie/.style={color=black, densely dotted, mark=x, mark options={solid, scale=1.1}},
  sCSIEP/.style={color=black, densely dashed, mark=triangle, mark options={solid, fill=white}},
  sCSIML/.style={color=black, solid, line width=0.5pt, mark=star, mark options={solid, scale=1.25}},
}

\newcommand{\resultlegend}{%
\begin{tikzpicture}
\begin{axis}[
  hide axis, scale only axis, width=1mm, height=1mm,
  xmin=0, xmax=1, ymin=0, ymax=1,
  legend columns=3,
  legend cell align=left,
  legend style={
    at={(0.5,0.5)}, anchor=center,
    font=\scriptsize, draw=black!35, line width=0.3pt,
    inner xsep=2pt, inner ysep=1pt,
    /tikz/every even column/.append style={column sep=5pt},
  },
  legend image post style={line width=0.7pt},
]
\addlegendimage{sLMMSE}\addlegendentry{LMMSE+EP}
\addlegendimage{sConv}\addlegendentry{Conv-JCED}
\addlegendimage{sSem, line width=1pt}\addlegendentry{Sem-JCED (proposed)}
\addlegendimage{sGenie}\addlegendentry{Genie-aided LMMSE}
\addlegendimage{sCSIEP}\addlegendentry{CSI-EP}
\addlegendimage{sCSIML, line width=0.5pt}\addlegendentry{CSI-ML}
\end{axis}
\end{tikzpicture}}

%% file: figures/tikz/fig_system.tex
\begin{tikzpicture}[
  font=\scriptsize,
  >={Stealth[length=3pt,width=2.4pt]},
  every path/.style={line width=0.45pt},
  blk/.style={draw, rounded corners=1.5pt, fill=black!5, minimum width=1.2cm,
    minimum height=0.6cm, align=center, inner sep=1pt},
  lab/.style={inner sep=1.5pt},
  cell/.style={draw=black!45, line width=0.2pt},
]
\def\antenna#1{%
  \draw (#1) -- ++(0,0.2);
  \draw ($(#1)+(0,0.2)$) -- ++(-0.085,0.14) -- ++(0.17,0) -- cycle;}
\def\yb{-2.55}

\foreach \d in {2,1}
  \draw[rounded corners=2pt, fill=white, draw=black!55]
    ($(-0.3,-0.48)+(\d*0.07,\d*0.07)$) rectangle ($(6.05,0.78)+(\d*0.07,\d*0.07)$);
\draw[rounded corners=2pt, fill=white] (-0.3,-0.48) rectangle (6.05,0.78);
\node[anchor=north west, lab] at (-0.3,0.78) {User $k\in\{1,\dots,K\}$};

\node (u) at (0,0) {$\mathbf u_k$};
\node[blk] (b2b) at (1.15,0) {Bytes\\to bits};
\node[blk] (qam) at (2.85,0) {QAM\\mapping};
\node[blk] (rgm) at (4.55,0) {RG\\mapping};
\draw[->] (u) -- (b2b);
\draw[->] (b2b) -- node[above, lab] {$\mathbf c_k$} (qam);
\draw[->] (qam) -- (rgm);
\draw[<-] (rgm.north) -- ++(0,0.3) node[left, lab, pos=0.75] {pilots};
\draw (rgm.east) -- (5.8,0);
\antenna{5.8,0}
\node[lab, anchor=north] at (5.52,-0.02) {$x_{s,c,k}$};

\node[draw, ellipse, fill=black!5, inner sep=0pt, align=center] (ch) at (7.4,-1.0)
  {Channel\\$\mathbf H_{s,c}$, $\mathbf n_{s,c}$};
\draw[->, line width=0.6pt] (6.25,0.45) to[out=0, in=100] (ch.north);
\draw[->, line width=0.6pt] (ch.south) to[out=-80, in=90] (8.05,\yb+0.47);

\draw[rounded corners=2pt, dashed, draw=black!60] (0.55,\yb-0.5) rectangle (8.55,\yb+0.58);
\node[anchor=south west, lab] at (0.55,\yb-0.5) {BS};
\foreach \x in {7.75,8.05,8.35} \antenna{\x,\yb+0.13}
\draw (7.75,\yb+0.13) -- (8.35,\yb+0.13);
\node[lab, anchor=north] at (8.05,\yb+0.13) {$N^{\rm r}$};
\node[blk] (ce) at (6.45,\yb) {Channel\\estimation};
\node[blk] (sd) at (4.45,\yb) {Symbol\\detection};
\node[blk] (dm) at (2.35,\yb) {Demapping};
\draw[->] (7.75,\yb+0.13) -- (7.75,\yb) -- node[above, lab] {$\mathbf y_{s,c}$} (ce.east);
\draw[->] (ce) -- node[above, lab] {$\hat{\mathbf H}_{s,c}$} (sd);
\draw[->] (sd) -- node[above, lab] {$\hat x_{s,c,k}$} (dm);
\draw[->] (dm.west) -- ++(-0.45,0) node[left, lab] {$\hat{\mathbf u}_k$};

\def\cw{0.13}\def\ch{0.105}
\begin{scope}[shift={(3.42,-1.66)}]
  \foreach \s in {1,...,14} \foreach \c in {1,...,8}
    \draw[cell, fill=white] ({(\s-1)*\cw},{(\c-1)*\ch}) rectangle ++(\cw,\ch);
  \foreach \s in {3,12} \foreach \c in {1,...,8} {
    \pgfmathtruncatemacro{\own}{mod(\c-1,4)==0 ? 1 : 0}
    \ifnum\own=1
      \draw[cell, fill=black!75] ({(\s-1)*\cw},{(\c-1)*\ch}) rectangle ++(\cw,\ch);
    \else
      \draw[cell, fill=black!18] ({(\s-1)*\cw},{(\c-1)*\ch}) rectangle ++(\cw,\ch);
    \fi}
  \draw[->, line width=0.35pt] (0,-0.07) -- ({14*\cw},-0.07) node[right, inner sep=1pt] {$s$};
  \draw[->, line width=0.35pt] (-0.07,0) -- (-0.07,{8*\ch}) node[above, inner sep=1pt] {$c$};
\end{scope}
\draw[densely dotted, draw=black!60] (rgm.south) -- (4.55,-0.82);
\begin{scope}[shift={(0.25,-1.0)}]
  \draw[cell, fill=black!75] (0,0) rectangle ++(\cw,\ch);
  \node[anchor=west, lab] at (0.17,0.05) {$\mathcal P_k$: pilots of user $k$};
  \draw[cell, fill=black!18] (0,-0.28) rectangle ++(\cw,\ch);
  \node[anchor=west, lab] at (0.17,-0.23) {$\mathcal P\setminus\mathcal P_k$: silent};
  \draw[cell, fill=white] (0,-0.56) rectangle ++(\cw,\ch);
  \node[anchor=west, lab] at (0.17,-0.51) {$\mathcal D$: data};
\end{scope}
\end{tikzpicture}%

%% file: figures/tikz/fig_receiver.tex
\begin{tikzpicture}[
  font=\scriptsize,
  >={Stealth[length=3pt,width=2.4pt]},
  every path/.style={line width=0.45pt},
  mod/.style={draw, rounded corners=1.5pt, fill=black!5, minimum width=1.55cm,
    minimum height=0.74cm, align=center, inner sep=1pt},
  eqs/.style={font=\tiny, text=black!60},
  lab/.style={inner sep=1.5pt},
  fb/.style={densely dashed},
]
\node[mod] (ce) at (1.25,0) {LMMSE\\estimator\\[-1pt]
  {\tiny\color{black!60}\eqref{eq:ce_pilot_obs}--\eqref{eq:ce_damping}}};
\node[mod] (ep) at (4.45,0) {EP\\detector\\[-1pt]
  {\tiny\color{black!60}\eqref{eq:whitening}--\eqref{eq:ep_damping}}};
\node[mod, fill=cSem!10, draw=cSem!70!black] (bt) at (7.25,0) {ByT5\\corrector\\[-1pt]
  {\tiny\color{black!60}\eqref{eq:detected_byte}--\eqref{eq:byteqam}}};
\node[draw, rounded corners=1.5pt, fill=white, align=center, inner sep=1.5pt] (cb) at (3.9,-1.42)
  {Combine\\[-1pt]{\tiny\color{black!60}\eqref{eq:semantic_tilted}, \eqref{eq:ep_estimation}}};

\draw (0.6,0.9) -- (4.45,0.9);
\draw[->] (1.25,0.9) -- (ce.north);
\draw[->] (4.45,0.9) -- (ep.north);
\node[lab, anchor=south west] at (0.6,0.9) {$\{\mathbf y_{s,c}\}_{(s,c)\in\mathcal G}$, $\sigma^2$};
\draw[<-] (ce.west) -- ++(-0.3,0) node[left, lab] {pilots};

\draw[->] (ce) -- node[above, lab] {$\hat{\mathbf H}_{s,c}^{(\ell)},\mathbf E_{s,c}^{(\ell)}$} (ep);
\draw[->] (ep) -- node[above, lab] {$p_{s,c,k}^{(\ell,T)}$} (bt);
\draw[->] (3.9,0|-ep.south) -- node[left, lab] {$q_{s,c,k}^{(\ell,T)}$} (cb.north);
\draw[->] (bt.south) -- node[right, lab] {$q_{s,c,k}^{\rm s,(\ell)}$} (7.25,-1.42) -- (cb.east);

\fill (4.95,-1.42) circle (0.9pt);
\draw[->, fb] (4.95,-1.42) -- (4.95,0|-ep.south);
\draw[->, fb] (cb.west) -- node[below, lab] {$\bar x_{s,c,k}^{(\ell)},\bar v_{s,c,k}^{(\ell)}$} (1.25,-1.42) -- (ce.south);

\end{tikzpicture}%

%% file: figures/tikz/fig_byt5.tex
\begin{tikzpicture}[
  font=\scriptsize,
  >={Stealth[length=3pt,width=2.4pt]},
  every path/.style={line width=0.45pt},
  byte/.style={draw=black!60, minimum width=0.38cm, minimum height=0.38cm,
    inner sep=0pt, text height=1.6ex, text depth=0.3ex},
  err/.style={byte, draw=cConv!85!black, text=cConv!85!black, line width=0.7pt},
  mod/.style={draw, rounded corners=1.5pt, minimum width=4.78cm,
    minimum height=0.44cm, align=center, inner sep=1pt},
  lab/.style={inner sep=1.5pt},
]
\foreach \c [count=\i] in {T,h,e,\textvisiblespace,c,h,q,n,n,e,l} {
  \ifnum\i=7 \node[err] (b\i) at ({0.44*(\i-1)},0) {\c};
  \else \node[byte] (b\i) at ({0.44*(\i-1)},0) {\c}; \fi}
\node[lab, anchor=south west] at (b1.north west) {Detected bytes $\hat{\mathbf u}_k^{(\ell)}$};
\node[lab, anchor=south, text=cConv!85!black] at (b7.north) {$j$};

\node[mod, fill=cSem!10, draw=cSem!70!black] (enc) at (2.2,-0.68)
  {ByT5-Small encoder $f_{\boldsymbol\theta^{\rm enc}}$};
\node[mod, fill=black!5] (cls) at (2.2,-1.3)
  {Classification head $f_{\boldsymbol\theta^{\rm cls}}$ and softmax};
\foreach \i in {1,...,11} \draw[->] (b\i.south) -- (b\i.south |- enc.north);
\draw[->] (enc) -- (cls);

\node[draw=black!45, rounded corners=1.5pt, minimum width=2.95cm,
  minimum height=1.9cm, anchor=south west] (pb) at (4.86,-1.52) {};
\node[lab, anchor=north west] at (pb.north west) {$q_{k,j}^{\rm b,(\ell)}(u)$ at position $j$};
\foreach \ch/\p/\r/\col in {a/0.93/0/cSem, q/0.04/1/cConv!85!black, e/0.01/2/black!60, o/0.01/3/black!60} {
  \node[anchor=base east, inner sep=0pt, text=\col] at (5.2,{-0.3-0.25*\r}) {\ch};
  \fill[\col] (5.27,{-0.3-0.25*\r}) rectangle ++({1.9*\p+0.02},0.16);
  \node[anchor=base west, inner sep=0pt, font=\tiny] at ({5.33+1.9*\p},{-0.3-0.25*\r}) {\p};
}
\node[inner sep=0pt] at (5.12,-1.24) {$\vdots$};
\draw[->] (cls.east) -- (cls.east -| pb.west);
\end{tikzpicture}%

%% file: figures/tikz/fig_convergence.tex
\def\convdata{figures/data/convergence_4x4_12dB.dat}
\resultlegend\par\vspace{2pt}
\subfloat[NMSE]{%
\begin{tikzpicture}
\begin{axis}[resultaxis, ymode=log,
  xlabel={Outer iteration $\ell$}, ylabel={NMSE},
  xmin=0, xmax=20, xtick={0,10,20},
  ymin=6e-4, ymax=1.3e-2]
\addplot[sLMMSE, mark repeat=5] table[x=iter, y=nmseLMMSE]{\convdata};
\addplot[sConv, mark repeat=3] table[x=iter, y=nmseConv]{\convdata};
\addplot[sSem, mark repeat=3] table[x=iter, y=nmseSem]{\convdata};
\addplot[sGenie, mark repeat=5] table[x=iter, y=nmseGenie]{\convdata};
\end{axis}
\end{tikzpicture}}\hspace{2pt}%
\subfloat[BER]{%
\begin{tikzpicture}
\begin{axis}[resultaxis, ymode=log,
  xlabel={Outer iteration $\ell$}, ylabel={BER},
  xmin=0, xmax=20, xtick={0,10,20},
  ymin=2e-3, ymax=1e-1]
\addplot[sLMMSE, mark repeat=5] table[x=iter, y=berLMMSEEP]{\convdata};
\addplot[sConv, mark repeat=3] table[x=iter, y=berConv]{\convdata};
\addplot[sSem, mark repeat=3] table[x=iter, y=berSem]{\convdata};
\addplot[sCSIEP, mark repeat=5] table[x=iter, y=berCSIEP]{\convdata};
\addplot[sCSIML, mark repeat=5] table[x=iter, y=berCSIML]{\convdata};
\end{axis}
\end{tikzpicture}}%

%% file: figures/tikz/fig_snr.tex
\def\snrfourfour{figures/data/snr_4x4.dat}
\def\snrfoureight{figures/data/snr_4x8.dat}
\resultlegend\par\vspace{2pt}
\subfloat[NMSE, $K=N^{\rm r}=4$]{%
\begin{tikzpicture}
\begin{axis}[resultaxis, ymode=log,
  xlabel={SNR (dB)}, ylabel={NMSE},
  xmin=4, xmax=16, xtick={4,8,12,16},
  ymin=2e-4, ymax=1e-1]
\addplot[sLMMSE] table[x=snr, y=nmseLMMSE]{\snrfourfour};
\addplot[sConv] table[x=snr, y=nmseConv]{\snrfourfour};
\addplot[sSem] table[x=snr, y=nmseSem]{\snrfourfour};
\addplot[sGenie] table[x=snr, y=nmseGenie]{\snrfourfour};
\end{axis}
\end{tikzpicture}}\hspace{2pt}%
\subfloat[BER, $K=N^{\rm r}=4$]{%
\begin{tikzpicture}
\begin{axis}[resultaxis, ymode=log,
  xlabel={SNR (dB)}, ylabel={BER},
  xmin=4, xmax=16, xtick={4,8,12,16},
  ymin=3e-4, ymax=3e-1]
\addplot[sLMMSE] table[x=snr, y=berLMMSEEP]{\snrfourfour};
\addplot[sConv] table[x=snr, y=berConv]{\snrfourfour};
\addplot[sSem] table[x=snr, y=berSem]{\snrfourfour};
\addplot[sCSIEP] table[x=snr, y=berCSIEP]{\snrfourfour};
\addplot[sCSIML] table[x=snr, y=berCSIML]{\snrfourfour};
\end{axis}
\end{tikzpicture}}\par\vspace{3pt}
\subfloat[NMSE, $K=4$, $N^{\rm r}=8$]{%
\begin{tikzpicture}
\begin{axis}[resultaxis, ymode=log,
  xlabel={SNR (dB)}, ylabel={NMSE},
  xmin=2, xmax=10, xtick={2,4,6,8,10},
  ymin=8e-4, ymax=1e-1]
\addplot[sLMMSE] table[x=snr, y=nmseLMMSE]{\snrfoureight};
\addplot[sConv] table[x=snr, y=nmseConv]{\snrfoureight};
\addplot[sSem] table[x=snr, y=nmseSem]{\snrfoureight};
\addplot[sGenie] table[x=snr, y=nmseGenie]{\snrfoureight};
\end{axis}
\end{tikzpicture}}\hspace{2pt}%
\subfloat[BER, $K=4$, $N^{\rm r}=8$]{%
\begin{tikzpicture}
\begin{axis}[resultaxis, ymode=log,
  xlabel={SNR (dB)}, ylabel={BER},
  xmin=2, xmax=10, xtick={2,4,6,8,10},
  ymin=1e-4, ymax=2e-1]
\addplot[sLMMSE] table[x=snr, y=berLMMSEEP]{\snrfoureight};
\addplot[sConv] table[x=snr, y=berConv]{\snrfoureight};
\addplot[sSem] table[x=snr, y=berSem]{\snrfoureight};
\addplot[sCSIEP] table[x=snr, y=berCSIEP]{\snrfoureight};
\addplot[sCSIML] table[x=snr, y=berCSIML]{\snrfoureight};
\end{axis}
\end{tikzpicture}}%